\documentclass[%
 reprint,
 amsmath,amssymb,
 aps,
 prl,
 longbibliography,
 lengthcheck,%
]{revtex4-1}

\usepackage{graphicx}
\usepackage{dcolumn}
\usepackage{bm}
\usepackage{hyperref}

\begin{document}

\preprint{APS/123-QED}

\title{Collective excitations in graphene in magnetic field}

\author{P. A. Andreev}
\email{andreevpa@physics.msu.ru}
 \affiliation{Department of General Physics, Physics Faculty, Moscow State
University, Moscow, Russian Federation.}

\date{\today}

\begin{abstract}
Collective excitations in graphene monolayer  are studied.
Equations describing collective properties of electrons in
graphene are obtained. The basic ideas of the method of
many-particle quantum hydrodynamics are used for the derivation.
As starting point of the derivation we use the Dirac equation for
massless electrons which is usually used for description of
electrons in graphene [D. E. Sheehy and J. Schmalian, Phys. Rev.
Lett. \textbf{99}, 226803 (2007)], where the Coulomb interaction
is taken into account. We study dispersion properties of
collective excitations by means derived here graphene quantum hydrodynamics
equations (GQHD). We consider graphene in the
external magnetic field which directed at an angle to the graphene
sample. We do it in a linear approximation of the GQHD equations. We
observe that the magnetic field directed perpendicular to the
graphene plane had no influence on dispersion of the collective
excitations. For the magnetic field directed at an angle to the
graphene we obtain dependence of wave dispersion on system
parameters: strength of magnetic field, wave vector, direction of
wave propagation relatively to the magnetic field.
\end{abstract}

\maketitle


\section{\label{sec:level1} I. Introduction}

Unusual properties of graphene conductivity ~\cite{Das Sarma RMP
11}, ~\cite{Peres RMP 10} have led to its widespread use in
semiconductor geterostructures. Important characteristic of three
dimensional (3D) and low dimensional conductors and semiconductors
is a dispersion  of collective excitations, particularly the plasma
waves or plasmons. Knowledge of the plasmon dispersion allows us to
analyze various processes in geterostructures. A model describing
the collective properties of graphene is constructed in this
paper. We use this model for definition of dispersion of the quantum
collective excitations in graphene at the presence of the external
magnetic field. The background of this research is the model presented in Ref. ~\cite{Sheehy PRL 07}. This model describes the
dynamic of electrons in graphene and reflects essential
characteristics of the conduction electrons in  graphene. The basic
equation describing microscopic dynamics of the conduction electrons
in graphene is the many-particle equation which can be presented
in the form analogous to the Schrodinger equation (or the Dirac equation)
~\cite{Sheehy PRL 07}
\begin{equation}\label{graph Schr eq general form}\imath\hbar\partial_{t}\psi=\hat{H}\psi,\end{equation}
where $\hat{H}$ is the Hamilton operator. In the absence of
external fields and interaction between particles the Hamiltonian
has form
$\hat{H}=\sum_{i=1}^{N}v_{F}\sigma_{i}^{\alpha}\hat{p}_{i}^{\alpha}$,
here $N$ is the number of particles in the system, $v_{F}$ is the
Fermi velocity of electrons in graphene, $\hat{p}_{i}^{\alpha}$ is
the momentum operator,
$\hat{p}_{i}^{\alpha}=-\imath\hbar\partial_{i}^{\alpha}$,
$\partial_{i}^{\alpha}$ is the derivative on coordinate of $i$-th
particle, $\sigma^{\alpha}_{i}$ are the Pauli matrixes. Described
model corresponds to the massless behavior of the conductivity
electrons in graphene ~\cite{Novoselov nature 05}. The Coulomb interaction between the electrons in graphene is considered in Ref.
~\cite{Sheehy PRL 07}.
Equation (\ref{graph Schr eq general form}) as the many-particle
Schrodinger equation is not always suitable for description of the
collective properties in many-particle systems. This problem
connects with the fact that equation (\ref{graph Schr eq general
form}) determines the wave
function in the 3N (2N, for two dimensional system) dimensional
configuration space, whereas the collective process realize in the 3D
(2D) physical space. Therefore, it is important to construct a
method in the 3D (2D) physical space ~\cite{Goldstein}. The method of
quantum hydrodynamics (QHD) solves the problem of transition from
configuration space to physical space. This method was suggested
for many-particle system and has been developed for a wide class
of physical systems. Method of the many-particle QHD (MPQHD) allows us
 derive equations for quantum observable values evolution in the 3D
(2D) physical space. Equations of continuity, momentum balance,
energy balance, momentum evolution (for particles with spin) and
polarization (for polarized particles) are appeared from the
Schrodinger equation. Most of them are analogous to classic
hydrodynamic equations. In connection with described above the
method is called "Quantum hydrodynamics". Method of the QHD is a
modern powerful method for studying of the collective properties in
systems of charged ~\cite{MaksimovTMP 1999}- ~\cite{Andreev PRB11},
and neutral ~\cite{Andreev PRA08}, ~\cite{Andreev arxiv 11 2}
particles (see also review ~\cite{Shukla RMP 11}). Unusual
properties of electrons of conductivity in graphene mapped in the
equation (\ref{graph Schr eq general form}) and mapped in the form
of the QHD equations for graphene (GQHD), which we present below. The equations of the GQHD in this paper has the extreme difference from the QHD
equations obtained previously  ~\cite{MaksimovTMP 1999} -
~\cite{Shukla RMP 11}. In context of using of the Dirac equation for
description of graphene electrons we note that in Ref.
~\cite{Asenjo PP 11} the QHD equation were derived from the bispinor Dirac
equation for massive particles. Obtained there equations were averaged on ensemble for
receiving of equations for many-particle system. Comparison of the QHD
equations for graphene with the usual QHD equations ~\cite{Andreev
PRB11}, ~\cite{Shukla RMP 11} we present below during the
derivation of the GQHD equations for graphene. As expected
unusual properties of electrons of conductivity in graphene lead
to exotic spectrum of elementary excitations. In 2D electron gas
(2DEG) the basic type of the collective excitations is the Langmuir
waves, the dispersion dependence of these waves are
$$\omega^{2}=\frac{2\pi e^{2}n_{0}k}{m}$$
where $e$, $m$ are the charge and mass of electrons ($e$ we
consider as algebraic quantity, $e=-|e|$, for electrons), $n_{0}$ is the 2D
concentration of electrons, $k$ is the absolute value of the wave
vector. For the graphene electrons in the absence of external
fields the dispersion equation is
$$\omega=kv_{F}.$$
So, we did not obtain the contribution of interaction in
dispersion, and, consequently, from our description the graphene
Langmuir frequency (see Ref. ~\cite{Das Sarma RMP 11} p.414) does
not follow.

The recent studies of excitations in graphene being in a magnetic field have employed the picture of single electron excitation and it's transition between energy levels, such as the Landau levels, which exist for single charged particle in an external field. Consideration of microscopic picture of excitations show that some of these excitations lead to the formation of collective excitations ~\cite{Roldan PRB 09}-~\cite{Goerbig RMP 11}.

If we want to study macroscopic collective excitations in the graphene, we do not need to consider one electron excitation. For described purpose we can use hydrodynamics or kinetics methods as it has been used for two dimensional electron gas, usual three dimensional carriers and semiconductors. These method get used to be used for getting of characteristics of the collective excitations in described objects, such as the Langmuir waves, the magneto-sound waves, the Bernstein modes, and etc. So, we can see that obtaining of the hydrodynamics description of graphene is the important task. The method of the many-particle QHD gives us possibility to derive GQHD equations, which make more rich a set of theoretical tools of graphene excitations studying.

Our paper is organized as follows. In Sec. II we present
derivation of the GQHD. In Sec. III we describe the method for
calculation of the dispersion of the collective excitations. In Sec.
IV dispersion of collective excitations for the graphene electrons in the
external magnetic field is studied. In Sec. V the brief
description of obtained results is presented.

\section{\label{sec:level mod} II. Construction of the model}

We derive basic equations by means of the MPQHD method
~\cite{MaksimovTMP 1999}, ~\cite{Andreev PRB11}. We use the
many-particle spinor massless Dirac equation ~\cite{Sheehy PRL 07},
~\cite{Novoselov nature 05}
\begin{equation}\label{graph Hamiltonian}\imath\hbar\partial_{t}\psi=\Biggl(\sum_{i}\biggl(v_{F}\sigma^{\alpha}D_{i}^{\alpha}+e_{i}\varphi_{i,ext}\biggr)+\sum_{i,j\neq i}\frac{1}{2}e_{i}e_{j}G_{ij}\Biggr)\psi\end{equation}
The following designations are used in the Hamiltonian (\ref{graph
Hamiltonian}):
$D_{i}^{\alpha}=-\imath\hbar\partial_{i}^{\alpha}-e_{i}A_{i,ext}^{\alpha}/c$,
 $\varphi_{i,ext}$, $A_{i,ext}^{\alpha}$ - is the potentials
of the external electromagnetic field,
$\textbf{E}_{i,ext}=-\nabla\varphi_{i,ext}-\partial_{t}\textbf{A}_{i,ext}$
is the electric field, $\textbf{B}_{i,ext}=curl\textbf{A}_{i,ext}$
is the magnetic field, quantities $e_{i}$, $m_{i}$-are the charge
and mass of particles, $\hbar$-is the Planck constant, and
$G_{ij}=1/r_{ij}$, - is the Green functions of the Coulomb
interaction. Replacing $-\imath\hbar\partial_{i}^{\alpha}$ by
$D_{i}^{\alpha}$ were used in Ref. ~\cite{Castro Neto RMP 09}
(see. p.127). It was made for account of external magnetic field In equation
(\ref{graph Hamiltonian}) the spinor wave function
$\psi=\psi(R,t)$ depend on 2N coordinates $R=[\textbf{r}_{1}, ...,
\textbf{r}_{N}]$ and time, where $\textbf{r}_{i}=[x_{i}, y_{i}]$
is the 2D coordinates of each particle. Potentials
$\varphi_{i,ext}=\varphi_{ext}(\textbf{r}_{i},t)$,
$A_{i,ext}^{\alpha}=A_{ext}^{\alpha}(\textbf{r}_{i},t)$ also
depend on 2D variables. This fact has deep consequences.
Potential part of electric field connected with the scalar
potential via space derivative:
$\textbf{E}_{i}=-\nabla_{i}\varphi_{i}$. Consequently in equation
(\ref{graph Hamiltonian}) there is no contribution of external
electric field directed perpendicular to the graphene plane $E_{z}$ (Contribution of $E_{z}$ might
appear via $\partial_{t}A_{z}$). Physically, there is no
limitation on attendance of $z$ projection of electric field and
it's action on graphene electrons. Especially if the graphene
sample is the part of the geterostructure or spin-field-effect
transistor ~\cite{Das Sarma RMP 11}, ~\cite{Britnell arxiv 11},
~\cite{ALD-FET}, ~\cite{Zutic RMP 04}, there exist the
contribution of external electric field in normal direction to the
graphene plane. The magnetic field vector to be
$$\textbf{B}=curl\textbf{A}=\textbf{e}_{x}(\partial_{y}A_{z}-\partial_{z}A_{y})$$
$$+\textbf{e}_{y}(\partial_{z}A_{x}-\partial_{x}A_{z})+\textbf{e}_{z}(\partial_{x}A_{y}-\partial_{y}A_{x}),$$
two component of the vector potential of the magnetic field
$A_{x}$, $A_{y}$are presented in Hamiltonian (\ref{graph
Hamiltonian}), and  they does not depend on coordinate $z$.
Therefore equation (\ref{graph Hamiltonian}) contain $z$ component
of the magnetic field only. In this paper we interested in action of the
external magnetic field directed at angle of graphene plane.
Therefore, we generalized GQHD equations including whole vector of
magnetic field $B\cdot e_{z}\rightarrow
\textbf{B}=[B_{x},B_{y},B_{z}]$.

We assumed that the Pauli matrices satisfy the following commutation
relation:
\begin{equation}\label{graph comm rel} [\sigma^{\alpha}_{i},\sigma^{\beta}_{j}]=2\imath\delta_{ij}\varepsilon^{\alpha\beta\gamma}\sigma^{\gamma}_{i}.\end{equation}

Graphene is the 2D structure and electrons of graphene are located
in the plane. As we describe above, in 2D case the electrons has
two coordinate $x$ and $y$, but spin of the electrons can be directed
in all direction, particularly, in direction of $z$ axes,
perpendicular to the graphene plane. This fact is accounted by
formula (\ref{graph comm rel}). Two projection of spin operator
are contained in the Hamiltonian (\ref{graph Hamiltonian}), these
are $\widehat{\sigma}_{x}$ and $\widehat{\sigma}_{y}$. Spin
operator projection on $z$ axis is appeared during derivation of the
GQHD equations due to commutation relation (\ref{graph comm rel}).

Now, we present the derivation of model of the electron collective
dynamics in graphene. As we mention above we describe the
derivation of the MPQHD equations from equation (\ref{graph
Hamiltonian}). The first step is we present the definition for density of
probability for the conduction electron system in the physical space.
\begin{equation}\label{graph def density}n(\textbf{r},t)=\sum_{s}\int dR\sum_{i}\delta(\textbf{r}-\textbf{r}_{i})\psi^{*}(R,t)\psi(R,t)\end{equation}
where $dR=\prod_{p=1}^{N}d\textbf{r}_{p}$.

 The quantity $n(\textbf{r},t)$ can be considered as the 2D concentration of the
conductivity electrons. For studying of the time evolution of the
concentration we differentiate the concentration (\ref{graph def
density}) with respect to time and use equation (\ref{graph
Hamiltonian}). In the result, we receive an equation which has the
form of the continuity equation:
\begin{equation}\label{graph cont usual}\partial_{t}n(\textbf{r},t)+\nabla \textbf{j}(\textbf{r},t)=0,\end{equation}
where
$$\textbf{j}(\textbf{r},t)=\sum_{s}\int dR\sum_{i}\delta(\textbf{r}-\textbf{r}_{i})\frac{1}{2}v_{F}\times$$
\begin{equation}\label{graph def of j}\times\Biggl(\psi^{*}_{s}(R,t)\biggl(\sigma^{\alpha}_{i}\psi\biggr)_{s}(R,t)+h.c.\Biggr)\end{equation}
and $\textbf{j}(\textbf{r},t)=v_{F}\textbf{S}(\textbf{r},t)$.

The quantity $\textbf{S}(\textbf{r},t)$ describes the spin density of
the particle system. Consequently, we have equation
\begin{equation}\label{graph cont with S}\partial_{t}n(\textbf{r},t)+v_{F}\nabla\textbf{S}(\textbf{r},t)=0.\end{equation}
Coordinate vector $\textbf{r}$ has only two component. Consequently, equation (\ref{graph cont with S}) contains two component of the spin density vector $\textbf{S}$, these are $S_{x}$ and $S_{y}$. Our next step in construction of the model of collective motion
is obtaining of an equation for spin evolution. For this aim we
differentiate the quantity $\textbf{S}(\textbf{r},t)$ with respect to time and use equation
(\ref{graph Hamiltonian}). Because we known the third component of spin density vector we can study the evolution of all components of this vector. Therefore, we derive equation of evolution for $\textbf{S}=[S_{x},S_{y},S_{z}]$. On this way we have equation of the spin
evolution:
\begin{equation}\label{graph evolut of S}\partial_{t}S^{\alpha}(\textbf{r},t)+v_{F}\partial^{\alpha}n(\textbf{r},t)=-\frac{2}{\hbar}\varepsilon^{\alpha\beta\gamma}J^{\beta\gamma}_{M}(\textbf{r},t)\end{equation} Here new
physical quantity is arisen: $J^{\alpha\beta}_{M}(\textbf{r},t)$.
The evident form of $J^{\alpha\beta}_{M}(\textbf{r},t)$ is
$$J^{\alpha\beta}_{M}(\textbf{r},t)= v_{F}\sum_{s}\int dR\sum_{i}\delta(\textbf{r}-\textbf{r}_{i})\frac{1}{2}\times$$
\begin{equation}\label{graph def of J mmm} \times\Biggl(\psi^{*}_{s}(R,t)\biggl(\sigma^{\alpha}_{i}D_{i}^{\beta}\psi\biggr)_{s}(R,t)+h.c.\Biggr).\end{equation}
This is the tensor of spin current. We study the 2D structures existing in 3D space. It leads to existence of $z$ component of
vectors for several physical quantities, as for the spin density
vector. The spin current $J^{\alpha\beta}_{M}(\textbf{r},t)$ is
defined via two-vectors (\ref{graph def of J mmm}). These are the
vector of spin $\sigma^{\alpha}$ and derivative vector operator
$D_{i}^{\beta}$, the last one is the vector with two components.
Therefore, we have the tensor $J^{\alpha\beta}_{M}(\textbf{r},t)$
whose first index take three value $x$, $y$, $z$, but the second
index take two value $x$, $y$. We have obtained two equation of the
GQHD, it is equations (\ref{graph cont with S}) and (\ref{graph
evolut of S}). These equations significantly vary from the first
two equations of the usual QHD ~\cite{Marklund
PRL07}-~\cite{Andreev PRB11}.
In the usual QHD as in the classic hydrodynamics the first equation is the
continuity equation where described the changing of number of
particles in vicinity of the point of the physical space in consequence
of the particles current. Instead of that we obtain the connection
of particles number changing and the spin density (\ref{graph cont
with S}).

We have obtained the equation (\ref{graph evolut of S}) instead of the momentum
balance equation (the Euler equation) in the usual QHD. The last one
accounts the influence of particles interaction as each other and
with the external fields. Equation (\ref{graph evolut of S}) does not
contain information about the interaction. Equation
(\ref{graph evolut of S}) differs from the usual equation of the spin
evolution (the Bloch equation) which contains the vector product of the spin density and
magnetic field whose leads to the spin evolution.

At this step we have no closed set of two equations (\ref{graph
def density}) and (\ref{graph evolut of S}). Thereto, received
equations do not contain information about interaction. Thus,
basic model, presented by equation (\ref{graph Hamiltonian}),
contains the Coulomb interaction between electrons. Therefore, we have the
series of equations which is an analogous to the BBGKI series ~\cite{Akhiezer}, ~\cite{Landau v10} in the classic physical kinetics. We can obtain next equation of
the series, namely equation of evolution for the tensor
$J^{\alpha\beta}_{M}(\textbf{r},t)$. This equation to be
$$\partial_{t}J^{\alpha\beta}_{M}(\textbf{r},t)+v_{F}\partial^{\alpha}J^{\beta}(\textbf{r},t)$$
$$-\hbar
v_{F}^{2}\varepsilon^{\alpha\gamma\delta}\partial^{\beta}\partial^{\gamma}S^{\delta}(\textbf{r},t)=-\frac{e}{c}v_{F}^{2}\varepsilon^{\alpha\beta\gamma}n(\textbf{r},t)B^{\gamma}(\textbf{r},t)$$
$$+\frac{2v_{F}^{2}}{\hbar}\varepsilon^{\alpha\mu\nu}\Pi^{\nu\mu\beta}(\textbf{r},t)+ev_{F}S^{\alpha}(\textbf{r},t)E^{\beta}(\textbf{r},t)$$
\begin{equation}\label{graph evol JM}-e^{2}v_{F}S^{\alpha}(\textbf{r},t)\partial^{\beta}\int d\textbf{r}'G(\textbf{r},\textbf{r}')n(\textbf{r}',t).\end{equation}
Equation (\ref{graph evol JM}) contains interaction and two new
quantities, these are $J^{\alpha}(\textbf{r},t)$ and
$\Pi^{\alpha\beta\gamma}$. They are
$$J^{\alpha}(\textbf{r},t)= v_{F}\sum_{s}\int dR\sum_{i}\delta(\textbf{r}-\textbf{r}_{i})\frac{1}{2}\times$$
\begin{equation}\label{graph def of
J}\times\Biggl(\psi^{*}_{s}(R,t)\biggl(D_{i}^{\alpha}\psi\biggr)_{s}(R,t)+h.c.\Biggr)\end{equation}
and
$$\Pi^{\alpha\beta\gamma}(\textbf{r},t)=\sum_{s}\int dR\sum_{i}\delta(\textbf{r}-\textbf{r}_{i})\frac{1}{2}\times$$
\begin{equation}\label{graph def of Pi}\times\Biggl(\psi^{*}_{s}(R,t)\biggl(\sigma^{\alpha}_{i}D_{i}^{\beta}D_{i}^{\gamma}\psi\biggr)_{s}(R,t)+h.c.\Biggr).\end{equation}

Equation (\ref{graph evol JM}) is the first equation in system of the GQHD equations which contain the magnetic field. In equation (\ref{graph evol JM}) we have made generalization and consider all three component of the magnetic field instead of $B_{z}$ which presented in the Hamiltonian (\ref{graph Hamiltonian}).

The second rank tensor
$\varepsilon^{\alpha\beta\gamma}B_{ext}^{\gamma}$ contains
strength of the external magnetic field and gives contribution to the
evolution of spin current. $B_{x}$ and $B_{y}$
do not include in equation (\ref{graph Hamiltonian}), but they can make influence on particles dynamic.

Quantity $\textbf{J}(\textbf{r},t)$ is an analog of the current of probability or the momentum
density in the usual QHD ~\cite{MaksimovTMP 1999}, ~\cite{Shukla RMP 11}, ~\cite{Andreev PRB11}. This is very
important physical quantity and we propose that our model must be
complete by one more equation - the momentum balance equation. For
derivation of the momentum balance equation, i.e. equation of
evolution of $\textbf{J}(\textbf{r},t)$, we apply the same method as used above. We
differentiate $\textbf{J}(\textbf{r},t)$ with respect to time and use the equation
(\ref{graph Hamiltonian}). In the result we have
$$ \partial_{t}J^{\alpha}(\textbf{r},t)+v_{F}\partial^{\beta}J^{\beta\alpha}_{M}=\frac{e v_{F}^{2}}{c}\varepsilon^{\alpha\beta\gamma}S^{\beta}B^{\gamma}$$
\begin{equation}\label{graph evol of J}+ev_{F}nE^{\alpha}-e^{2}v_{F}n\partial^{\alpha}\int d\textbf{r}'G(\textbf{r},\textbf{r}')n(\textbf{r}',t).\end{equation}
Equation (\ref{graph evol of J}) is very similar to the Euler
equation in the usual QHD. But instead of the Lorentz force we have vector
product of the spin density and the magnetic field.

 Equation (\ref{graph evol of J}), just as
equation (\ref{graph evol JM}), contains interaction between
particles. In fact equation (\ref{graph evol of J}) is an analog
of the Euler equation in the usual QHD ~\cite{MaksimovTMP 1999},
~\cite{Shukla RMP 11}, ~\cite{Andreev PRB11}. Obtained set of
equations (\ref{graph cont with S}), (\ref{graph evolut of S}),
(\ref{graph evol JM}) and (\ref{graph evol of J}) is not close
because the set of equations contains quantity
$\Pi^{\alpha\beta\gamma}(\textbf{r},t)$. We can obtain
equation for $\Pi^{\alpha\beta\gamma}(\textbf{r},t)$, but in this
way we will obtain new physical quantities. We need to make a
closed set of the equations (\ref{graph cont with S}), (\ref{graph
evolut of S}), (\ref{graph evol JM}) and (\ref{graph evol of J}),
expressing $\Pi^{\alpha\beta\gamma}(\textbf{r},t)$ via physical
quantities included in these equations. For closing the set of
equations (\ref{graph cont with S}), (\ref{graph evolut of S}),
(\ref{graph evol JM}) and (\ref{graph evol of J}) we use method of
approximate calculations of the hydrodynamic variables developed in
~\cite{MaksimovTMP 1999}, ~\cite{Andreev PRA08}. Following the
paper ~\cite{MaksimovTMP 1999}, ~\cite{Andreev PRA08} we obtain
$$\Pi^{\alpha\beta\gamma}(\textbf{r},t)=S^{\alpha}(\textbf{r},t)\frac{J^{\beta}(\textbf{r},t)J^{\gamma}(\textbf{r},t)}{n^{2}(\textbf{r},t)}$$
\begin{equation}\label{graph appr evid form Pi}+\varrho^{\alpha\beta\gamma}(\textbf{r},t)-\hbar^{2}S^{\alpha}(\textbf{r},t)\frac{\partial^{\beta}\partial^{\gamma}\sqrt{n(\textbf{r},t)}}{\sqrt{n(\textbf{r},t)}}.\end{equation}
where $\varrho^{\alpha\beta\gamma}(\textbf{r},t)$ describes the contribution of the thermal motion, so, it is an analog of the tensor of kinetic pressure in the usual hydrodynamics. In this paper we do not account the thermal motion. Consequently we suggest that $\varrho^{\alpha\beta\gamma}(\textbf{r},t)$ is equal to zero. The last term in formula (\ref{graph appr evid form Pi}) is an analog of the quantum Bohm potential (see for example ~\cite{Andreev PRB11}).

Now, we have a closed system of the equations (\ref{graph cont with
S}), (\ref{graph evolut of S}), (\ref{graph evol JM}), (\ref{graph
evol of J}) and (\ref{graph appr evid form Pi}). Using the GQHD equations (\ref{graph cont
with S}), (\ref{graph evolut of S}), (\ref{graph evol JM}),
(\ref{graph evol of J}) and (\ref{graph appr evid form Pi}) we can study the collective excitations and
they properties, i.e. dispersion dependence and increments of
instabilities.

Developing method allows us to derive the energy balance
equation which give us possibility to study the influence of the
temperature on the graphene dynamics.

\subsection{Self-consistent field approximation: discussion}

In last terms of equations (\ref{graph evol JM}) and (\ref{graph
evol of J}) we have used the self-consistent field approximation. Here
we present general expressions for two-particle functions appeared
in equations (\ref{graph evol JM}) and (\ref{graph evol of J}),
and also describe the meaning of the self-consistent approximation.
This approximation
was suggested by A. A. Vlasov in 1938  ~\cite{Vlasov JETP 38} for many-particle system of charged particles.

\begin{figure}
\includegraphics[width=8cm,angle=0]{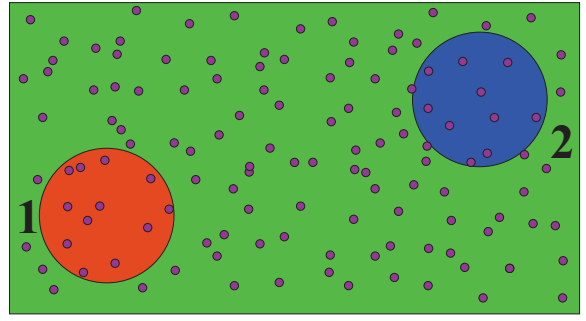}
\caption{\label{selfCint01} The figure presents the picture of the
self-consistent interaction in the system of charged particles.}
\end{figure}

The Fig. \ref{selfCint01} presents the picture of the self-consistent
interaction between charges particles. Total charge of the region 2
interacts with the total charge of the region 1. Changing the
extreme point of a radius vector (or shifting the region 2) we can
scan whole space. In this way we obtain action of the external charges
on the region 1. Changing position of the region 1 and repeating described
operation we obtain action surrounding charges on each region of
space. This is a picture of the self-consistent interaction in fixed
moment of time and this picture governs an evolution of particles
in system. This picture of interaction is typical for the classic
physics, where we need to obtain smooth functions describing the
collective motion. For that is necessary to average at physically
infinitesimal volume (sketched circle). In the quantum mechanics, the concentration, the spin density, the current density, etc, are defined
via wave function and we can consider described picture on
interaction of separate particle instead of the space regions.

In general case the last terms in the equations (\ref{graph evol JM}) and (\ref{graph evol of J}) contain the two-particle function which in the self-consistent field approximation expressed via one particle functions and in this approximation we obtain a closed set of equations. In the equations (\ref{graph evol JM}) and (\ref{graph evol of J}) the following two-particle functions were appeared. Two-particle concentration
$$n_{2}(\textbf{r},\textbf{r}',t)=\sum_{s}\int dR\times$$
\begin{equation}\label{graph}\times\sum_{i,j\neq i}\delta(\textbf{r}-\textbf{r}_{i})\delta(\textbf{r}'-\textbf{r}_{j})\psi^{+}_{s}(R,t)\psi_{s}(R,t) \end{equation}
is the average of the product of the concentration operators $\sum_{i=1}^{N}\delta(\textbf{r}-\textbf{r}_{i})$, where $N$ is the number of particles in the system. The quantity
$$j_{2}^{\alpha}(\textbf{r},\textbf{r}',t)=\sum_{s}\int dR$$
\begin{equation}\label{graph}\times\sum_{i,j\neq i}\delta(\textbf{r}-\textbf{r}_{i})\delta(\textbf{r}'-\textbf{r}_{j})v_{F}\psi^{+}_{s}(R,t)\biggl(\sigma^{\alpha}_{i}\psi\biggr)_{s}(R,t)\end{equation}
is the two-particle function of the concentration and the spin density $\sum_{i=1}^{N}\sigma_{i}^{\alpha}\delta(\textbf{r}-\textbf{r}_{i})$. Technically, the self-consistent field approximation corresponds to the factoring of the two-particle functions in product of one-particle ones:
\begin{equation}\label{graph}n_{2}(\textbf{r},\textbf{r}',t)\rightarrow n(\textbf{r},t)n(\textbf{r}',t)\end{equation}
and
\begin{equation}\label{graph}j_{2}^{\alpha}(\textbf{r},\textbf{r}',t)\rightarrow v_{F}S^{\alpha}(\textbf{r},t)n(\textbf{r}',t).\end{equation}
It was used at derivation of the equations (\ref{graph evol JM}) and
(\ref{graph evol of J}). This approximation is suitable at
consideration of the long-range interaction, particularly for the Coulomb
interaction.

If we interest in more detail description of interaction we need
to consider many-particle correlations. For example, for the
two-particle concentration the correlation to be
$$g(\textbf{r},\textbf{r}',t)=n_{2}(\textbf{r},\textbf{r}',t)- n(\textbf{r},t)n(\textbf{r}',t),$$
where $g(\textbf{r},\textbf{r}',t)$ includes a quantum correlation
caused by exchange interaction. A method of correlation
calculation was developed in Ref.s ~\cite{MaksimovTMP 1999},
~\cite{Andreev PRA08} and ~\cite{MaksimovTMP 2001(2)}. This method
also can be used for the graphene description, and  for it's further and more
detailed studying.

\section{\label{sec:level meth of calc} III. Method of calculation of wave dispersion}

We consider the small perturbation of the equilibrium state like
$$\begin{array}{ccc}n=n_{0}+\delta n,& S^{\alpha}=S_{0}^{\alpha}+\delta S^{\alpha},& \textbf{S}_{0}\parallel \textbf{B}_{0}\end{array}$$
\begin{equation}\label{di BEC equlib state BEC}\begin{array}{ccc}& & J^{\alpha}=0+\delta J^{\alpha}, J_{M}^{\alpha\beta}=0+\delta J_{M}^{\alpha\beta},\end{array}\end{equation}
where $\textbf{B}_{0}$ is the external magnetic field,
$\textbf{B}_{0}=[B_{0x},0,B_{0z}]$.

Substituting these relations into system of the GQHD equations
(\ref{graph cont with S}), (\ref{graph evolut of S}), (\ref{graph
evol JM}), (\ref{graph evol of J}) and (\ref{graph appr evid form
Pi}) \textit{and} neglecting by nonlinear terms, we obtain a
system of linear equations with constant coefficients.

Passing to the following representation for small perturbations
$\delta f$
$$\delta f =f(\omega, \textbf{k}) exp(-\imath\omega t+\imath \textbf{k}\textbf{r}) $$
yields the homogeneous system of algebraic equations

\begin{equation}\label{graph lin eq cont}-\imath\omega \delta n+\imath v_{F}\textbf{k}\delta \textbf{S}=0, \end{equation}
\begin{equation}\label{graph lin eq evol spin}-\imath\omega\delta S^{\alpha}+\imath v_{F}k^{\alpha}\delta n=-\frac{2}{\hbar}\varepsilon^{\alpha\beta\gamma}\delta J_{M}^{\beta\gamma},\end{equation}
$$-\imath\omega\delta J^{\alpha}+\imath v_{F}k^{\beta}\delta J^{\beta\alpha}=\frac{ev_{F}^{2}}{c}\varepsilon^{\alpha\beta\gamma}B_{0}^{\gamma}\delta S^{\beta}$$
\begin{equation}\label{graph lin eq evol current}+\imath k^{\alpha}e^{2}v_{F}n_{0}(2\pi/k)\delta n\end{equation}
and
$$-\imath\omega\delta J_{M}^{\alpha\beta}+\imath v_{F}k^{\alpha}\delta J^{\beta}+\hbar v_{F}^{2}k^{\beta}k^{\gamma}\varepsilon^{\alpha\gamma\delta}\delta S^{\delta}$$
$$=-\frac{e}{c}v_{F}^{2}\varepsilon^{\alpha\beta\gamma}B_{0}^{\gamma}\delta n+v_{F}^{2}\hbar\varepsilon^{\alpha\mu\nu}\frac{1}{n_{0}}S_{0}^{\nu}k^{\mu}k^{\beta}\delta
n$$
\begin{equation}\label{graph lin eq evol spin current}+\imath e^{2}v_{F}S_{0}^{\alpha}k^{\beta}(2\pi/k)\delta n.\end{equation}

We assume that the spin density magnitude has a nonzero value.
Expressing all quantities entering in the set of equations in
terms of the spin density, we get the equation
$$\Lambda^{\alpha\beta}(\omega,\textbf{k})\cdot S^{\beta}(\omega,\textbf{k})=0,$$
where
$$\Lambda^{\alpha\beta}(\omega,\textbf{k})=\omega^{2}\delta^{\alpha\beta}-v_{F}^{2}k^{\alpha}k^{\beta}$$
$$+2\frac{e}{c}\frac{1}{\hbar\omega}\varepsilon^{\alpha\gamma\mu}\varepsilon^{\gamma\delta\nu}B_{0}^{\nu}k^{\beta}k^{\delta}k^{\mu}\frac{v_{F}^{5}}{\omega^{2}-v_{F}^{2}k^{2}}$$
$$-2\frac{e}{\hbar c}\varepsilon^{\alpha\gamma\mu}\varepsilon^{\beta\gamma\nu}B_{0}^{\nu}k^{\mu}\frac{\omega v_{F}^{3}}{\omega^{2}-v_{F}^{2}k^{2}}+2\varepsilon^{\alpha\gamma\mu}\varepsilon^{\beta\delta\mu}v_{F}^{2}k^{\gamma}k^{\delta}$$
$$+2\frac{e}{c}\varepsilon^{\alpha\gamma\mu}\varepsilon^{\gamma\mu\nu}B_{0}^{\nu}\frac{1}{\hbar\omega}v_{F}^{3}k^{\beta}+2\varepsilon^{\alpha\gamma\delta}\varepsilon^{\delta\mu\nu}v_{F}^{3}k^{\beta}k^{\gamma}k^{\mu}S_{0}^{\nu}\frac{1}{n_{0}\omega}$$
\begin{equation}\label{graph disp matrix GENERAL}+4\pi\imath e^{2}\varepsilon^{\alpha\gamma\delta}k^{\beta}k^{\gamma}S_{0}^{\delta}\frac{1}{\hbar\omega k}.\end{equation}
Contribution of the Coulomb interaction is presented by the last term
in the dispersion matrix $\Lambda^{\alpha\beta}$.

In the absence of external magnetic field the dispersion matrix $\Lambda^{\alpha\beta}$ can be presented as
$$\Lambda^{\alpha\beta}(\omega,\textbf{k})=\omega^{2}\delta^{\alpha\beta}-v_{F}^{2}k^{\alpha}k^{\beta}$$
$$+2\varepsilon^{\alpha\gamma\mu}\varepsilon^{\beta\delta\mu}v_{F}^{2}k^{\gamma}k^{\delta}.$$

Dispersion equation to be
\begin{equation}\label{graph disp eq 1} det\widehat{\Lambda}(\omega,\textbf{k})=0.\end{equation}
This equation can be represent in the form
\begin{equation}\label{graph disp eq 2} (\Lambda_{xx}\Lambda_{yy}-\Lambda_{xy}\Lambda_{yx})\Lambda_{zz}=0,\end{equation}
and splits into two
\begin{equation}\label{graph disp eq 2a} \Lambda_{xx}\Lambda_{yy}-\Lambda_{xy}\Lambda_{yx}=0,\end{equation}
and
\begin{equation}\label{graph disp eq 2b} \Lambda_{zz}=0.\end{equation}

\section{\label{sec:level GR in magn field} IV. Graphene in the magnetic field}

In this section we consider dispersion properties of waves in
graphene placed in the external magnetic field which parallel or
perpendicular to the $XY$ plane, we suppose that graphene is
located in the $XY$ plane.

We begin this chapter with the consideration of equation
\begin{equation}\label{graph disp eq with Mg field Lxy in gen}\Lambda_{xx}\Lambda_{yy}-\Lambda_{xy}\Lambda_{yx}=0.\end{equation}

For the magnetic field perpendicular to the graphene plane
$\textbf{B}_{0}=B_{z}\textbf{e}_{z}$ from (\ref{graph disp eq with
Mg field Lxy in gen}) we have
\begin{equation}\label{graph disp solution with Mg field Lxy Bz}\omega=v_{F}k.\end{equation}
It is the same result as in the absence of the external field.

\begin{figure}
\includegraphics[width=8cm,angle=0]{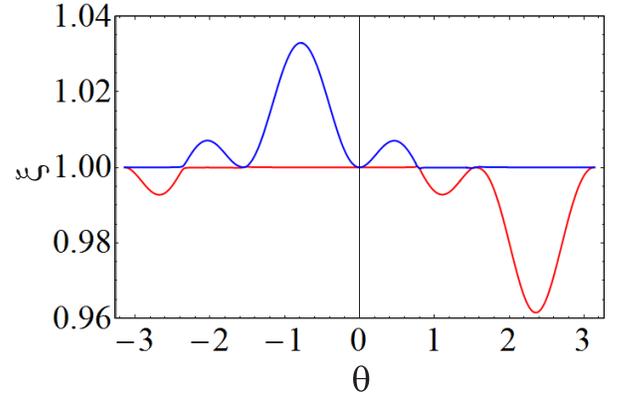}
\caption{\label{LxyTwoInOne} The figure presents the dependence of
the reduced frequency $\xi$ (\ref{graph red frec def}), appearing from equation (\ref{graph disp eq with Mg field Lxy in gen}) in presence of the in-plane magnetic field $B_{0x}$, on the
angle $\theta$, which is the angle between axes Ox and direction
of the wave propagation. Dimensionless parameters of the system
are choose to be equal to the following quantities  $\kappa=-10^{-6}$, $\chi=0.1$, and $\mu=10^{-3}$. Blue curve (upper) lying at $\xi\geq 1$ corresponds to one of solutions of equation (\ref{graph disp eq with Mg field Lxy in gen}) and red curve (lowest) lying at $\xi\leq 1$ described the second solution.}
\end{figure}

Suggesting what $S_{0x}\sim B_{0x}$ we can admit that equation (\ref{graph disp eq with Mg field
Lxy in gen}) contains terms proportional to $B_{0x}$ and $B_{0x}^{2}$. Considering the limit of approximately small in-plane magnetic field $B_{0x}$ we can neglect by the terms proportional to the $B_{0x}^{2}$. In the result we find an equation which is the equation of eight degree of $\omega$. However, at the considering values of the parameters of the system we get that this equation has two solutions only. These solutions reveal strong nonmonotone dependence of the collective excitation frequency $\omega$ on the angle between direction of wave propagation and the direction of the in-plane magnetic field (\ref{LxyTwoInOne}).   

From equation (\ref{graph disp eq with Mg field Lxy in gen}) we find that in described approximation
$\xi$ depends on three dimensionless parameters
$$\kappa=\frac{eB_{0x}}{c\hbar k^{2}}<0,$$
$$\mu=\frac{e^{2}S_{0z}}{\hbar v_{F}k^{2}},$$ 
$$\chi=S_{0x}/n_{0},$$
and angle of wave propagation $\theta$. The angle $\theta$ is defined as
$\cos\theta=k_{x}/k$ and $\sin\theta=k_{y}/k$.

Let's consider equation $\Lambda_{zz}=0$. Dispersion equation has following explicit form
\begin{equation}\label{graph disp eq with Mg field Lzz evident}\omega^{2}+2v_{F}^{2}k^{2}-2\frac{e}{\hbar c}\frac{\omega v_{F}^{3}\textbf{k}\textbf{B}_{0}}{\omega^{2}-v_{F}^{2}k^{2}}=0\end{equation}
or in dimensionless form to be
\begin{equation}\label{graph disp eq with Mg field gen}\xi^{2}+2-\frac{2\alpha\xi}{\xi^{2}-1}=0\end{equation}
where
\begin{equation}\label{graph red frec def}\xi=\frac{\omega}{kv_{F}}\end{equation}
and
\begin{equation}\label{graph param of ext field}\alpha=\frac{e}{\hbar c}\frac{\textbf{k}\textbf{B}}{k^{3}}=-\frac{|e|}{\hbar c}\frac{B_{0}}{k^{2}}\cos\theta.\end{equation}
$\xi$ describe the frequency of the wave divided by $kv_{F}$, $\alpha$ present the contribution of the external magnetic field which parallel to the graphene plane and $\theta$ is the angle between the magnetic field and the direction of the excitation propagation.

In the absence of the external field or if the magnetic field directed perpendicular to the graphene plane from (\ref{graph disp eq with Mg field gen}) we have
$$\xi^{2}+2=0,$$
thus the equation $\Lambda_{zz}=0$ has no wave solution.

\begin{figure}
\includegraphics[width=8cm,angle=0]{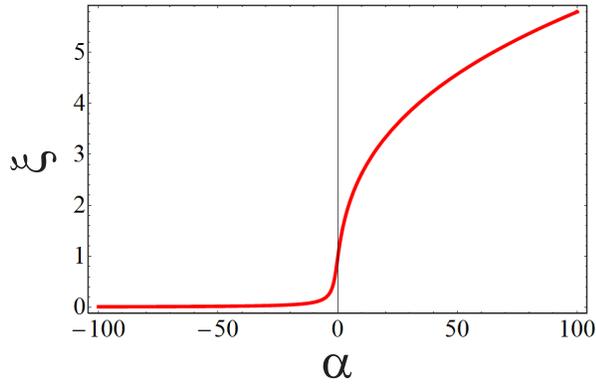}
\caption{\label{GRW1} The figure presents the dependence of the
reduced frequency $\xi$ (\ref{graph red frec def}) on parameter
$\alpha$ (\ref{graph param of ext field}).}
\end{figure}

Here we consider equation (\ref{graph disp eq with Mg field gen})
in the case when the magnetic field parallel to the graphene
plane, then the equation (\ref{graph disp eq with Mg field gen})
we represent in the form
\begin{equation}\label{graph disp eq with Mg field}\xi^{4}+\xi^{2}-2\alpha\xi-2=0.\end{equation}
The dependence of $\xi$ on $\alpha$ is presented Fig. \ref{GRW1}.
This solution shows no instabilities. We obtain this solution from
the equation
\begin{equation}\label{graph magnitudes connection}\Lambda_{zx}S_{x}+\Lambda_{zy}S_{y}+\Lambda_{zz}S_{z}=0,\end{equation}
since $\Lambda_{zz}=0$ and $S_{z}\neq 0$ we can note that obtained
wave solution contains wave of spin, where amplitude of spin
density directed perpendicular to the graphene plane and, thus, to
the direction of the wave propagation. We need to pay attention to
the fact that the Coulomb interaction contained in the basic
equations and the dispersion matrix does not influence on found
dispersion dependence.

\section{\label{sec:level concl}V. Conclusion}

We investigated  the influence of the external static uniform
magnetic fields on the dispersion properties of linear waves of
the graphene electrons. We supposed that the external field
directed at an angle to the plane where graphene is located. We
paid attention to the particular cases when the angle between the
magnetic field direction and the graphene plane equal to $0$ or
$\pi/2$. In the absence of external fields the dispersion
dependence of the collective excitation has form $\omega=v_{F}k$
(see formula (\ref{graph disp solution with Mg field Lxy Bz})). If
the magnetic field perpendicular to the graphene plane there is no
changes in the dispersion in compare with the case of graphene in
the absence of the external field. We obtained dispersion
dependence for the case magnetic field parallel to the graphene
plane and studied dependence of the frequency of the collective
excitations on strength of the external magnetic field and angle
between the magnetic field and the direction of excitation
propagation. We also showed that in the presence of an in-plane 
magnetic field $B_{0x}$, along with $B_{0z}$, solution $\omega=v_{F}k$ splits on two.

For studying of described problem we derived system of the QHD
equations for electrons in graphene. For this derivation we used
the method of the MPQHD. Obtained GQHD equations consist of four
equations: equations of concentration evolution, spin evolution,
current evolution and spin current evolution. In this equation we
made generalization and included in the equations the contribution
of $B_{x}$, $B_{y}$ along with the $B_{z}$.


\section{\label{sec:level1} Acknowledgments}
The author thanks Professor L. S. Kuz'menkov for fruitful discussions.



\end{document}